\documentclass[twocolumn, nofootinbib]{revtex4}
\usepackage{amsmath} \usepackage{graphicx} \usepackage{dcolumn}
\usepackage{bm} \usepackage{epsf} \usepackage{amssymb}

\begin{document}

\sloppy

\newcommand{\prtl}{\partial}
\newcommand{\la}{\left\langle}
\newcommand{\ra}{\right\rangle}
\newcommand{\dla}{\la \! \! \! \la}
\newcommand{\dra}{\ra \! \! \! \ra}
\newcommand{\we}{\widetilde}
\newcommand{\ws}{\we{s}}
\newcommand{\wF}{\we{F}}
\newcommand{\sins}{{\mbox{\scriptsize ins}}}
\newcommand{\smfp}{{\mbox{\scriptsize mfp}}}
\newcommand{\sfull}{{\mbox{\scriptsize full}}}
\newcommand{\smp}{{\mbox{\scriptsize mp}}}
\newcommand{\sLZ}{{\mbox{\scriptsize LZ}}}
\newcommand{\sph}{{\mbox{\scriptsize ph}}}
\newcommand{\sinhom}{{\mbox{\scriptsize inhom}}}
\newcommand{\sneigh}{{\mbox{\scriptsize neigh}}}
\newcommand{\srlxn}{{\mbox{\scriptsize rlxn}}}
\newcommand{\svibr}{{\mbox{\scriptsize vibr}}}
\newcommand{\smicro}{{\mbox{\scriptsize micro}}}
\newcommand{\smax}{{\mbox{\scriptsize max}}}
\newcommand{\seq}{{\mbox{\scriptsize eq}}}
\newcommand{\sstr}{{\mbox{\scriptsize str}}}
\newcommand{\teq}{{\mbox{\tiny eq}}}
\newcommand{\sinn}{{\mbox{\scriptsize in}}}
\newcommand{\tin}{{\mbox{\tiny in}}}
\newcommand{\scr}{{\mbox{\scriptsize cr}}}
\newcommand{\sscr}{{\mbox{\scriptsize scr}}}
\newcommand{\sL}{{\mbox{\scriptsize L}}}
\newcommand{\sTS}{{\mbox{\scriptsize TS}}}
\newcommand{\stheor}{{\mbox{\scriptsize theor}}}
\newcommand{\sGS}{{\mbox{\scriptsize GS}}}
\newcommand{\sNMT}{{\mbox{\scriptsize NMT}}}
\newcommand{\sRFOT}{{\mbox{\scriptsize RFOT}}}
\newcommand{\tRFOT}{{\mbox{\tiny RFOT}}}
\newcommand{\sbulk}{{\mbox{\scriptsize bulk}}}
\newcommand{\tbulk}{{\mbox{\tiny bulk}}}
\newcommand{\tDC}{{\mbox{\tiny DC}}}
\newcommand{\sauto}{{\mbox{\scriptsize auto}}}
\newcommand{\tauto}{{\mbox{\tiny auto}}}
\newcommand{\sescape}{{\mbox{\scriptsize escape}}}
\newcommand{\sth}{{\mbox{\scriptsize th}}}
\newcommand{\svib}{{\mbox{\scriptsize vib}}}
\newcommand{\sT}{{\mbox{\scriptsize T}}}
\newcommand{\sTLS}{{\mbox{\scriptsize TLS}}}
\newcommand{\sd}{{\mbox{\scriptsize d}}}
\newcommand{\sext}{{\mbox{\scriptsize ext}}}
\newcommand{\scav}{{\mbox{\scriptsize cav}}}
\newcommand{\bmu}{\bm \mu}
\newcommand{\bE}{\bm E}
\newcommand{\bD}{\bm D}
\newcommand{\bd}{\bm d}
\newcommand{\br}{\bm r}
\newcommand{\bj}{\bm j}
\newcommand{\bi}{\bm i}
\newcommand{\bnabla}{\bm \nabla}
\newcommand{\smol}{{\mbox{\scriptsize mol}}}
\def\Xint#1{\mathchoice
   {\XXint\displaystyle\textstyle{#1}}%
   {\XXint\textstyle\scriptstyle{#1}}%
   {\XXint\scriptstyle\scriptscriptstyle{#1}}%
   {\XXint\scriptscriptstyle\scriptscriptstyle{#1}}%
   \!\int}
\def\XXint#1#2#3{{\setbox0=\hbox{$#1{#2#3}{\int}$}
     \vcenter{\hbox{$#2#3$}}\kern-.5\wd0}}
\def\ddashint{\Xint=}
\def\dashint{\Xint-}
\title{Charge and momentum transfer in supercooled melts: \\ Why
  should their relaxation times differ?}

\author{Vassiliy Lubchenko} \affiliation{Department of Chemistry,
  University of Houston, Houston, TX 77204-5003}


\begin{abstract}

  The steady state values of the viscosity and the intrinsic
  ionic-conductivity of quenched melts are computed, in terms of
  independently measurable quantities. The frequency dependence of the
  ac dielectric response is estimated.  The discrepancy between the
  corresponding characteristic relaxation times is only apparent; it
  does not imply distinct mechanisms, but stems from the intrinsic
  barrier distribution for $\alpha$-relaxation in supercooled fluids
  and glasses.  This type of intrinsic ``decoupling'' is argued not to
  exceed four orders in magnitude, for known glassformers. We explain
  the origin of the discrepancy between the stretching exponent
  $\beta$, as extracted from $\epsilon(\omega)$ and the dielectric
  modulus data.  The actual width of the barrier distribution always
  grows with lowering the temperature. The contrary is an artifact of
  the large contribution of the dc-conductivity component to the
  modulus data.  The methodology allows one to single out other
  contributions to the conductivity, as in ``superionic'' liquids or
  when charge carriers are delocalized, implying that in those
  systems, charge transfer does not require structural
  reconfiguration.

\end{abstract}

\date{\today}

\maketitle

\section{Introduction}

Molecular motions in deeply supercooled melts and glasses are
cooperative so that transporting a single molecule requires concurrent
rearrangement of up to several hundreds of surrounding molecules. Such
high degree of cooperativity results in high barriers even for the
smallest scale molecular translations. These high barriers underly the
slow, activated dynamics in deeply supercooled melts and the emergence
of a mechanically stable aperiodic lattice, if a melt is quenched
sufficiently rapidly.  The Random First Order Transition (RFOT)
methodology, developed by Wolynes and coworkers, provides a
constructive microscopic picture of the structural rearrangements in
supercooled melts and quenched glasses. The RFOT has quantitatively
explained or predicted the signature phenomena accompanying the glass
transition, including the connection between the thermodynamic and
kinetic anomalies \cite{KTW, XW, XWbeta}, the length scale of the
cooperative rearrangements \cite{XW}, deviations from Stokes-Einstein
hydrodynamics \cite{XWhydro}, aging \cite{LW_aging}, the low
temperature anomalies \cite{LW, LW_BP, LW_RMP}, and more. (See
\cite{LW_ARPC} for a recent review.)

Perhaps the most dramatic experimental signature of the glass
transition is the rapid super-Arrhenius growth of the relaxation times
with lowering the temperature, from about a picosecond, near the
melting point $T_m$, to as long as hours, at the glass transition
temperature $T_g$. These relaxation times are deduced via several
distinct experimental methodologies and all display an extraordinarily
broad dynamical range.  Nevertheless, making detailed comparisons
between those distinct methodologies has required additional
phenomenological assumptions. Mysteriously, these comparisons show a
significant degree of mismatch, sometimes by several orders of
magnitude. For example, the phenomenological ``conductivity relaxation
time'' $\tau_\sigma$ \cite{Macedo1972}, is consistently shorter than
the mechanical relaxation time $\tau_s$, especially at lower
temperatures. The apparent time scale separation varies wildly from
system to system: For instance in molten nitrates, it is about four
orders of magnitude at $T_g$ \cite{Howell1974}, while in silver
containing superionic melts, the $\tau_s/\tau_\sigma$ ratio becomes as
large as $10^{11}$ \cite{McLinAngell}, i.e. almost as much as the
whole dynamical range accessible to the melt!  This disparity
suggested that the mechanical relaxation and the electrical
conductivity in these systems were in fact due to distinct mechanisms:
At higher temperatures, the time scale separation is small so that the
two processes strongly affect each other, or ``mix'', while at lower
temperatures, the processes become increasingly ``decoupled''
\cite{Angell}. At such low temperatures, the mechanical relaxation
occurs via the aforementioned, activated concerted events, also called
the primary, or $\alpha$-relaxation.  Other processes that seem to
decouple from the mechanical relaxation include nuclear spin
relaxation, rotational diffusion, and the diffusion of small
probes. (For reviews, see \cite{Angell, Ngai, CiceroneEdiger})

Here we focus on two specific transport phenomena: low-frequency
momentum transfer, i.e. the viscous response, and the ionic conduction
in supercooled melts. Notwithstanding the complications needed to
analyze the electrical modulus data \cite{Elliott1994, Roling1998,
  ColeTombari, Sidebottom1995, Moynihan1994, Dyre1991, Doi1988}, the
mismatch between the typical relaxation times, corresponding to the
two types of transport, is clearly present. Furthermore, in the case
of superionic compounds, one may show (see below) that conduction
occurs without distorting the liquid's structure beyond the typical
vibrational displacements.  This is much less obvious for compounds
where the ionic motions are ``decoupled'' from the bulk structural
relaxation by four orders of magnitude or less, the latter dynamic
range comparable to the breadth of the $\alpha$-peaks in dielectric
dispersion in insulating melts near $T_g$ (see
e.g. \cite{LSBL}). Accounting for the distribution width is essential
here because the viscosity and conductivity are distinct, in fact
exactly reciprocal types of response: In momentum transfer, the
velocity gradient is the source, and the passed-on rate-of-force is
the response.  In charge transfer, the force on the ion is the source,
while the arising velocity field is the response. Consistent with this
general notion, in computing the viscosity, we will average the
relaxation time, with respect to local inhomogeneities, while the
intrinsic ionic conductivity will be determined by the average {\em
  rate} of $\alpha$-relaxation. Because of the mentioned, extremely
broad distribution of structural relaxation times $\tau$, the quantity
$ \la \tau \ra \la \tau^{-1} \ra $ may reach several orders of
magnitude, and so an apparent decoupling is indeed expected; no
additional microscopic mechanisms need to be invoked. 

The microscopic calculation and comparison, of the viscosity and the
ionic conductivity, are thus the main focus of this article. The two
quantities are computed, in terms of the barrier distribution and
other measurable material properties, in Sections \ref{visc} and
\ref{cond} respectively. To perform comparisons with experiment and
assess the upper limit on the ``inherent decoupling'' between the two
phenomena, we will discuss the barrier distribution in some detail, in
Section \ref{Barrier}. We will find that indeed, the degree of
decoupling should increase with the width of the barrier distribution,
and hence at lower temperatures, as demonstrated by the RFOT
methodology \cite{XWbeta}. We will assess the deviation between
$\alpha$-relaxation times, as deduced from viscosity, ionic
conductance, and the maximum in $\epsilon''(\omega)$.  Further, we
will exemplify potential ambiguities in using the dielectric modulus
formalism in estimating the relaxation time distribution. The latter
techique has suggested that for some substances, the distribution
width in fact decreases with lowering the temperature, in conflict
with the present results and the correlation between the stretching
exponent $\beta$ and temperature, predicted earlier by the RFOT theory
\cite{XWbeta}. We will see that the conflict is artificial and results
from the large contribution of the ac-component to the modulus data,
consistent with earlier, phenomenological arguments \cite{Johari1988,
  Roling1998}.

\section{Viscosity}
\label{visc}

The key microscopic notion behind the RFOT methodology is that, {\em
  regardless of the detailed interparticle potentials}, local
aperiodic arrangements in classical condensates become meta{\em
  stable} below a certain temperature $T_A$ (or above a certain
density) \cite{dens_F1, dens_F2}. Chemical detail and molecular
structure affect the value of $T_A$, and the viscosity of the
fluid. If the viscosity is high enough, one may cool the liquid so
that it becomes locally trapped in metastable minima, while avoiding
the nucleation of a periodic crystal, which would have been the lowest
free energy state. Another system-dependent quantity is the size $a$
of the elemental structural unit in the metastable liquid, or
``bead'': the length $a$ plays the role of the lattice spacing in the
aperiodic structure, and is indeed quite analogous to the size of the
unit cell in an oxide crystal, or it may correspond to the size of a
rigid monomer or side chain in a polymer. The size $a$ characterizes
the range of the local chemical order that sets in during a crossover,
at a temperature $T_\scr$, from collision dominated transport to
activated dynamics \cite{LW_soft}. The temperature $T_\scr$ is related
to the meanfield temperature $T_A$ but is always smaller. The bead
size $a$ may be unambiguously determined from the fusion entropy of
the corresponding crystal, when the latter entropy is known
\cite{LW_soft}, or else can be computed from the fragility $D$ using
the universal relationship between the latter and the heat capacity
jump at $T_g$: $D = 32/\Delta c_p$, as derived in RFOT \cite{XW,
  StevensonW}. Alternatively, if the configuration entropy can be
reliably estimated, one may use the RFOT-derived relation for the
configurational entropy per bead $s_c \sim .8 k_B$ \cite{XW}, which is
somewhat sensitive to the barrier-softening effects though
\cite{LW_soft}.

\begin{figure}[t]
  \includegraphics[width= .95 \columnwidth]{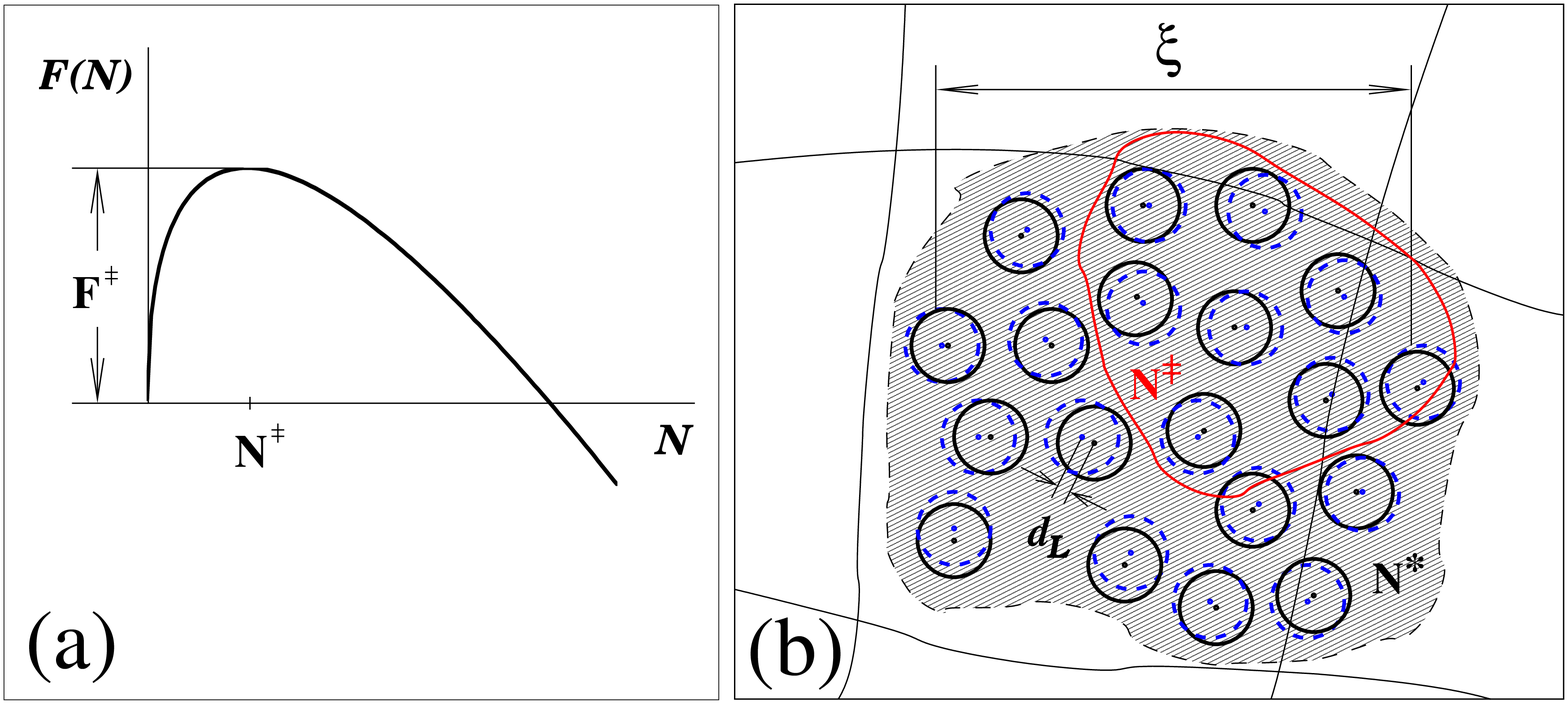}
  \caption{\label{xiFN} {\bf (a):} Typical nucleation profile of one
    aperiodic lattice, within another, in deeply supercooled liquids.
    $N \equiv (4\pi/3) (r/a)^3$, $N^\ddagger$ is the typical
    transition state size: $(dF/dN)_{N^\ddagger} = 0$. $\xi$ is the
    volumetrically defined cooperative length: $N^* \equiv (\xi/a)^3$,
    where $F(N^*) = 0$. {\bf (b):} Cartoon of a structural
    rearrangement. The shown magnitude of $\xi$ corresponds to a
    temperature near $T_g$ on 1 hour scale. The two sets of circles -
    solid and dashed ones - denote two alternative structural
    states. $d_L \simeq a/10$, or the ``Lindemann length'', is the
    typical bead displacement during a transition.}
\end{figure} 
  
Once locally metastable, the liquid may reconfigure but in an
activated fashion, i.e. by nucleating a new aperiodic structure within
the present one.  Such activated events occur, on average, once per
typical $\alpha$-relaxation time $\tau$, per region of size $\xi$.
The nucleus grows in a sequence of individual, nearly-barrierless bead
moves of length $d_L \simeq a/10$ \cite{dens_F1, L_Lindemann} and time
$\tau_\smicro \simeq 1$ ps \cite{L_Lindemann}. The overall sequence of
elemental moves typically corresponds to the following activation
profile, see Fig.\ref{xiFN}(a):
\begin{equation} \label{F(N)} F(N) = \gamma \sqrt{N} - T s_c N,
\end{equation}
where $N$ is the size of a reconfigured region. The ``surface term''
$\gamma \sqrt{N}$ is the mismatch penalty for creating one aperiodic
structure within another. $s_c$ is the excess, ``configurational''
entropy of the liquid per bead, hence the entropic, bulk term $(-T s_c
N)$, which drives the transition and reflects the multiplicity of
possible aperiodic arrangements in a region of size $N$. The maximum
of the profile:
\begin{equation} \label{Fsc} F^\ddagger = \mbox{max}\{F(N)\} =
  \frac{\gamma^2}{4 s_c T}
\end{equation}
is achieved at $N^\ddagger= N^*/4$, where $F(N^*)=0$, so that the
typical relaxation time is
\begin{equation} \label{tau} \tau = \tau_\smicro e^{F^\ddagger/k_B T}
  \equiv \tau_\smicro e^{D T_K/(T - T_K)}.
\end{equation}
This formula works well at $\tau > 1$ nsec or so.  The form on the
r.h.s. is the Vogel-Fulcher law, derived in the RFOT.

The end result of a cooperative, activated event is a reconfigured
region of size $\xi$, where each of the $N^* \equiv (\xi/a)^3$ beads
has moved the Lindemann length $d_L$, or so, see Fig.\ref{xiFN}(b).
Both $\xi$ and the nucleation critical size, $r^\ddagger =
\xi/4^{1/3}$, increase with lowering the temperature, roughly as
$r^\ddagger \propto \xi \propto 1/(T - T_K)^{2/3}$ \cite{KTW, XW,
  LW_soft}. Here, $T_K$ is the so called ideal glass transition
temperature, where the excess liquid entropy $s_c$, extrapolated below
$T_g$, would vanish \cite{Kauzmann}.  At $T_g$, $\xi$ is still quite
modest, only about six beads across \cite{XW, LW}.  Activated
transport becomes dominant below the temperature $T_\scr$, such that
$r^\ddagger(T_\scr) = a$, which corresponds, apparently universally,
to $\tau/\tau_\smicro \simeq 10^3$, or viscosities on the order of 10
Ps \cite{LW_soft}. At times shorter than $10^3 \tau_\smicro$, one may
then speak of a local aperiodic lattice on length scales of the
cooperativity size $\xi$, since the slow structural reconfigurations
have now time-scale separated from the vibrations
\cite{LW_soft}. Because of the local nature of structural relaxations,
one speaks of dynamic heterogeneity, or a ``mosaic'' of cooperative
rearrangments \cite{XW}. The heterogeneity is two-fold: On the one
hand, a {\em local} rearrangment implies that the surrounding
structure is static during the transition, up to vibrations. On the
other hand, because of the spatial and temporal variation in the local
density of states (and hence variations in $s_c$), local
reconfigurations are generally subject to somewhat different barriers
in different regions \cite{XW} (see Eq.(\ref{Fsc}) and also Section
\ref{Barrier}).

Computation of the viscosity in such a dynamically heterogeneous
environment may be done in two steps: First compute the viscosity in a
medium with a homogeneous relaxation time, call it $\tau'$, and then
average out with respect to the true distribution of the relaxation
times. This procedure is valid in view of the equivalence of time and
ensemble average.  When the relaxation rate is strictly spatially
homogeneous, one may formally define a diffusion constant for an
individual bead: $D' = d_L^2/6 \tau'$, since a bead moves the
Lindemann length, once per time $\tau'$, on average. Note that since a
bead's movements, as embodied in $d_L$ and $\tau'$, are dictated by
its cage, this is an example of ``slaved'' motion, to borrow
Frauenfelder's adjective for conformational changes of a protein
encased in a stiff solvent \cite{FFMP, LWF}.  One may associate, by
detailed balance, a low-frequency drag coefficient to that diffusion
constant: $\zeta' = k_B T/D'$. Such dissipative response implies
irreversible momentum exchange between a chosen particle and its
homogeneous (!)  surrounding, hence a Stokes' viscosity $\eta' =
\zeta'/(6 \pi a/2)$, where $a/2$ is used for the radius of the region
carved out in the liquid by a single bead.  Averaging with respect to
$\tau'$ yields for the steady state viscosity of the actual
heterogeneous liquid:
\begin{equation} \label{eta1} \eta = \frac{2 k_B T}{\pi a d_L^2} \la
  \tau \ra,
\end{equation}
where we have removed the prime at $\tau$, implying averaging with
respect to the actual barrier distribution.  The equation above can be
rewritten as
\begin{equation} \eta = (a/d_L)^2 \frac{2 k_B T}{\pi a^3} \la \tau \ra
  \simeq 60 \: \frac{k_B T_g}{a^3} \la \tau \ra,
\end{equation}
since the Lindemann ratio, $d_L/a \simeq 0.1$ has been argued to
change at most by 10\% between $T_m$ and $T_g$ \cite{L_Lindemann}.

Another instructive way to present Eq.(\ref{eta1}) is to note that the
Lindemann length is nearly equal to the typical amplitude of
high-frequency vibrations: $d_L \simeq d_\svibr$, within 5\% or so,
see Fig.3 of Ref.\cite{L_Lindemann}. The vibrational amplitude is
fixed by the equipartition theorem, since per bead: $K_\infty a^3
(d_\svibr/a)^2 = k_B T$, where $K_\infty$ is the high-frequency
elastic constant of the aperiodic lattice. One gets, as a result, a
Maxwell-type expression:
\begin{equation} \label{eta3} \eta \sim K_\infty \la \tau \ra.
\end{equation}
The last equation provides an easy way to see that the estimates in
Eqs.(\ref{eta1})-(\ref{eta3}) agree well with the experiment: Judging
from the sound speed in glasses \cite{FreemanAnderson,
  BerretMeissner}, the typical high frequency elastic modulus is about
$10^{9} - 10^{10}$ Pa, i.e. comparable but somewhat less than those of
crystals. The range of relaxation times $10^{-12} - 10^{4}$ sec,
implies $10^{-3} - 10^{13}$ Pa$\cdot$sec for the viscosity, as is
indeed observed. Alternatively, one may obtain these figures by
substituting a typical $a \sim 3$\AA ~(see \cite{LW_soft, StevensonW}
for specific estimates of bead sizes/densities).

Finally, the exploited equivalence between the time and ensemble
averages implies that crystallization has not begun during the
experiment, of course. The latter possibility adds uncertainty into
viscosity measurements, as the presence of crystallites would greatly
broaden the dynamic range of local relaxations owing to relatively
slow crystal nucleation events and the slow hydrodynamics near the
crystallites.  Similarly, long chain motions in polymeric melts would
also introduce additional long time scales into the problem. Our
derivation does not apply to those situations. We note that optical
transparency, which is often used as an indicator of no-crystallinity,
does not ensure that crystallites - hundreds of nanometers across or
smaller - are absent. Therefore a rigorous experimental study should,
in the least, check whether performing viscoelastic measurements has
enhanced the crystallization of the sample. Ideally, X-ray diffraction
should be monitored in the course of viscosity measurements.

\section{Ionic Conductivity}
\label{cond} 

\begin{figure}[t]
\includegraphics[width=.95\columnwidth]{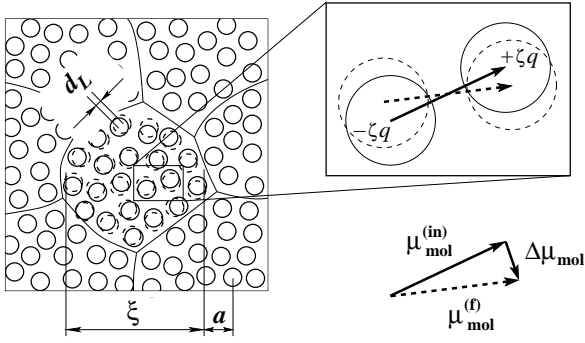}
\caption{\label{domain_dipole} Shown on the left is a fragment of the
  mosaic of cooperatively reconfiguring regions in the supercooled
  liquid.  Expanded portion shows how rotation of a bond leads to
  generating an elemental dipole change during a transition, where the
  partial charges on the two beads are $\pm \zeta q$.}
\end{figure}

In any supercooled melt, whether regarded ionic or not, the beads
carry an additional charge, relative to the corresponding crystal,
because of the lack of crystalline symmetry. As a result, each
structural reconfiguration is characterized by a transition-induced
electric dipole, see Fig.\ref{domain_dipole}.

The latter was estimated to be about a Debye or so, for most molecular
substances \cite{LSWdipole}. This value comes about as we may break up
the whole domain into $N^*/2$ pairs, where is pair has an elemental
dipole $\bmu_i$, $i = 1 \ldots N^*/2$: $\la |\bmu_i| \ra = \zeta e
q$. Here $q$ is the elementary charge: $q = 10^0 e$, and $\zeta < 1$
characterizes the excess charge. This quantity $\zeta$ is usually
small reflecting small deviations from the crystalline symmetry, in
the case of molecular crystals, or reflecting the weak interaction in
Van der Waals systems.  Alternatively, the overall density of
charged/polar beads may be low.  Ionic melts, by the very meaning of
the term, are distinct from molecular/Van der Waals systems in that
nearly {\em all} beads are strongly charged, implying $\zeta \sim 1$.
To be more specific, the conclusions of this article will be
exemplified with an often studied mixture of 40\%
Ca(NO$_3$)$_2$-60\%KNO$_3$ (``CKN''), $T_g \simeq 330$K.
 
During a transition, each dipole turns by an angle $(d_L/a)$. The
total transition dipole: 
\begin{equation} \bmu_T = \sum_{j}^{N^*/2} \left[ \bmu_j^{(f)} -
    \bmu_j^{(in)} \right]
\end{equation}
scales as $\sqrt{N^*}$ because of the random orientation of the
elemental dipoles \cite{LSWdipole}:
\begin{equation}
  \mu_T \simeq \zeta (q a) [(\xi/a)^3/2]^{1/2} (d_L/a).
\label{mu_Tqual}
\end{equation} 
When the dipole density is uniform, every transition results in a
local arrangement equally representative of the liquid structure.  In
other words, structural transitions do not modify the overall pattern
of the immediate coordination shell.  Transitions lead to a local
ionic currents: $\bi' = \bmu_T'/\tau'$, per region of volume $\xi^3$.
In the presence of an electric field, the net current density:
\begin{equation}
  \bj  = \la \bj' \ra = \frac{1}{\xi^3} \la \frac{\bmu_T}{\tau} \ra,
\end{equation}
is non-zero because the dipole moment at the transition state is
correlated with the overall transition dipole moment.  The latter can
be shown using Wolynes' library construction of liquid states
\cite{LW_aging}.  Repeating that argument, but in the presence of
electric field $\bE$, yields for the typical free energy profile for
structural reconfiguration in steady state:
\begin{equation} \label{F(N)E} F(N) = \gamma \sqrt{N} - \bmu_T(N)
  \bE_c - T s_c N,
\end{equation}
where $N$ is the size of the rearranged region and
\begin{equation}
  \bmu_T(N) = \sum_j^{N/2} \left[ \mu_j^{(f)} - \mu_j^{(in)} \right] 
\end{equation}
is the overall dipole change in that region. The subscript ``$c$'' in
$\bE_c$ signifies that the latter is a cavity field. The field
dependent term $|\bmu_T(N) \bE_c| \ll k_B T$ is overwhelmingly smaller
than the other terms in Eq.(\ref{F(N)E}). As a result, the transition
state dipole moment is not field-induced, but, again, is ``slaved'' to
the lattice. More formally, one may use the argument from
Ref.\cite{LW} showing that the density of structural states at the
reconfiguration bottle-neck is of the order $1/T_g$, implying the
field will not affect the specific sequence of elemental moves, but
will affect the dynamics merely by shifting the energies along
structurally dictated sequences of moves. Thus in the lowest order in
$\bE_c$, $\tau^{-1}(\bE_c) = \tau^{-1}(\bE_c=0)(1+ \bmu_T^\ddagger
\bE_c/k_B T)$, yielding
\begin{equation} \bj = \la \frac{\bmu_T (\bmu_T^\ddagger \bE_c)}{\tau}
  \ra \frac{1}{k_B T},
\end{equation}
where $\bmu_T^\ddagger \equiv \bmu_T(N^\ddagger)$ is the transition
dipole moment at the zero-field transition state. The cavity field,
see e.g. \cite{TitulaerDeutch}, is related to the external field
$E^{(\infty)}$ by
\begin{equation} E_c(\omega) = E^{(\infty)}(\omega) \frac{3
    \epsilon_b(\omega)}{2 \epsilon_b(\omega)+1},
\end{equation}
where $\epsilon_b(\omega)$ is the dielectric constant of the
surrounding bulk. Since a steady current is implied in the derivation,
(the imaginary part of) $\epsilon_b(\omega)$ diverges at zero
frequency, implying $E_c = E^{(\infty)} (3/2)$.  One thus obtains for
the ionic conductivity tensor:
\begin{equation} \sigma_{ij} = \la \frac{(\bmu_T)_i
    (\bmu_T^\ddagger)_j}{\tau} \ra \frac{3}{2 k_B T \xi^3}.
\end{equation}
Bearing in mind that $N^\ddagger = N^*/4 = (\xi/a)^3/4$, and that the
liquid is isotropic, on average: $\la (\Delta \bmu)_i (\Delta \bmu)_j
\ra = \delta_{ij} (\Delta \mu)^2/3$, one finally has:
\begin{equation} \label{sij} \sigma_{ij} = \delta_{ij} \, \frac{\Delta
    \mu_\smol^2}{8 \, a^3 k_B T} \la \frac{1}{\tau} \ra,
\end{equation}
where
\begin{equation} \Delta \mu_\smol^2 \equiv \la \left[ \mu_j^{(f)} -
    \mu_j^{(in)} \right]^2 \ra
\end{equation}
is the average elemental dipole change squared. Finally note that in
covalently networked materials, where dipole assignment may be
ambiguous, one may still estimate local dipole changes using the known
piezoelectric properties of the corresponding crystal, see
\cite{LSWdipole} for details.

The derivation above does not apply to systems where the dipole
density is significatly non-uniform. For instance, glycerol has one
polar, OH group per non-polar, aliphatic group, implying the liquid is
non-homogeneous, dipole moment wise, on the $\alpha$-relaxation time
scale. In such systems, the premise that structural rearrangements
result in equally representative configurations of the liquid does not
hold. In the glycerol example, ionic conduction would imply breaking
OH or CH bonds. In CKN, on the other hand, the overall bond pattern,
around any atom, does not change significantly during a transition,
even though individual bonds distort by the Lindemann length, as
mentioned.  Eq.(\ref{sij}) thus places the absolute upper limit on the
{\em intrinsic} ionic conductivity of a melt. By ``intrinsic'' we mean
that the computed currents are always present in the fluid and result
from the intrinsic activated transport: Local bond pattern does not
change significantly in the course of an individual activated event,
but only in the course of many consecutive events, since during an
individual event, the molecular displacements barely exceed typical
vibrational displacements. Conversely, if a system displays a higher
conductivity than prescribed by Eq.(\ref{sij}), one may conclude that
the ion motion does not require structural reconfiguration. Here, the
bond pattern actually changes, however these are not scaffold bonds of
the aperiodic lattice comprising the fluid (or glass). (More on this
below.)

To simplify comparison of Eq.(\ref{sij}) with experiment, let us
express the combination of the bead charge $\zeta q$ and size $a$, in
Eq.(\ref{sij}), through the {\em finite} frequency dielectric
response, a measurable quantity in principle (see below). The latter
is the response of a rearranging region in the absence of bulk
current, i.e. with a fixed environment, up to vibrations. It is
convenient to choose such regions at volume $\xi^3$, so that each
region has two structural states available, within thermal reach from
each other, separated by a barrier sampled from the actual barrier
distribution in the liquid.  If the two states, ``1'' and ``2'' are
characterized by dipole moments $\bmu_1$ and $\bmu_2$ respectively,
the expectation value of the dipole moment of the region is $\mu =
(\bmu_1 + \bmu_2)/2 + \Delta \bmu (p_2 - p_1)/2$, where $\Delta \bmu
\equiv (\bmu_2 - \bmu_1)$; $p_1$ and $p_2 = (1 - p_1)$ are the
probabities to occupy state 1 and 2 respectively. The relative
population $(p_2 - p_1)$ depends on the field via $\delta \ln(p_2/p_1)
= \Delta \bmu \bE_c/k_B T$. At realistic field strengths, i.e.
$|\Delta \bmu \bE_c/k_B T| \ll 1$, one has for the field-induced shift
of the relative population: $\delta (p_2 - p_1) \simeq 2 p_1 p_2
(\Delta \bmu \bE_c)$. Similarly to the preceding argument, $\la
(\Delta \bmu)_i (\Delta \bmu)_j \ra = (N^*/2) \delta_{ij} (\Delta
\mu_\smol)^2/3$. Further, since we have frozen the structural
transitions in the surrounding region, in estimating the cavity field,
one must use $\epsilon(\omega)$ with the $\alpha$-relaxation
contribution subtracted. This does not introduce much ambiguity
because in most ionic substances, the dielectric constant even at very
high frequences is significantly larger than unity. In CKN, for
instance, $\epsilon'(\infty) \simeq 7$ \cite{Howell1974}, allowing us
to write as before: $E_c \simeq E^{(\infty)}(3/2)$. One thus obtains,
in a standard fashion, for the frequency dependent dielectric constant
in the absence of macroscopic current:
\begin{equation} \label{epsmu} \epsilon_\sins(\omega)-\epsilon_\infty
  = 4 \pi \la p_1 p_2 \ra \frac{\Delta \mu_\smol^2}{4 \, a^3 k_B T}
  \la \frac{1}{1 - i \omega \tau} \ra,
\end{equation}
where the label ``ins'' signifies the absence of dc conductivity.

In the presence of steady current, the full response per domain is the
sum of the steady current from Eq.(\ref{sij}) and the ac current from
Eq.(\ref{epsmu}). The addition of the dc contribution, $i
\sigma/\omega$, to the full dielectric response will increase the
absolute value of $\epsilon_b(\omega)$.  This means that
Eq.(\ref{epsmu}) should work even better. One thus gets for the full
dielectric response of a conducting substance:
\begin{equation} \label{epsfull} \epsilon(\omega) - \epsilon_\infty =
  4 \pi \la p_1 p_2 \ra \frac{\Delta \mu_\smol^2}{4 \, a^3 k_B T} \la
  \frac{1}{1 - i \omega \tau} \ra+ i \frac{\sigma}{\omega},
\end{equation}
where $\sigma$ is the dc conductivity from Eq.(\ref{sij}).

One needs to know the distribution of the transition energies $E$ to
estimate the quantity $\la p_1 p_2 \ra \equiv \la 1/4 \cosh^2(E/2k_B
T) \ra$. Since $\xi$ is the smallest possible size of a rearranging
unit, these rearrangments correspond to the elementary excitations in
the system. We thus conclude, based on equipartition, that the typical
value of $E$ is roughly $k_B T$, implying that $\la p_1 p_2 \ra$ is
close to its maximum value of one quarter but is likely smaller by
another factor of two or so. Assuming then, for the sake of argument
that $\la p_1 p_2 \ra = 1/8$, one gets $\frac{\Delta \mu_\smol^2}{ a^3
  k_B T} \simeq 2 (\epsilon'_0-\epsilon_\infty)$, within an order of
magnitude. By Eq.(\ref{sij}), this implies a Maxwell-like relation
between the dc conductivity and the real part of the dielectric
response:
\begin{equation} \label{Maxwell_cond} \sigma \sim
  (\epsilon'_0-\epsilon_\infty) \la \frac{1}{\tau} \ra,
\end{equation}
with an important distinction, though, that here one averages the {\em
  inverse} relaxation time.  The $(\epsilon'_0-\epsilon_\infty)$
difference in CKN, to be concrete, is about $(20 - 7) = 13$
\cite{Pimenov, Howell1974}. This implies, by Eq.(\ref{epsmu}) and
CKN's $T_g \simeq 330$K \cite{Pimenov}, that $\zeta q \simeq 3e$, at
$a = 3$\AA, a reasonable value for the bead charge. Note that $a =
3$\AA ~is consistent with CKN's $\Delta c_p \simeq .12$ cal/g K
\cite{AngellTorell} and the mentioned $D = 32/\Delta c_p$ \cite{XW}.

Now, substituting CKN's $(\epsilon'_0-\epsilon_\infty)$ into
Eq.(\ref{sij}) yields for the conductivity $\sigma \simeq 3 \cdot
10^{-10} \la \tau^{-1} \ra$ (Ohm m)$^{-1}$sec.  Naively replacing $\la
\tau^{-1} \ra$ with $1/\la \tau \ra$ would imply, at the glass
transition, where $\la \tau \ra \sim 10^{2} - 10^{3}$ sec, a
conductivity of the order $10^{-13} - 10^{-14}$ (Ohm cm)$^{-1}$, which
is three to four orders of magnitude below the observed value
\cite{Pimenov, Howell1974}.  Note that the value $\la \tau^{-1} \ra
\la \tau \ra = 10^3 - 10^4$ is just the magnitude of decoupling
observed in CKN near $T_g$ \cite{Howell1974, Angell}, and is in fact
expected for a fragile substance such as CKN is, as we will argue in
the following.

\section{Barrier distribution and the Decoupling}
\label{Barrier}

Relaxation barriers in supercooled liquids are distributed because the
local density of states is non-uniform, leading to variations in the
local value of the configurational entropy and hence the RFOT-derived
barrier from Eq.(\ref{Fsc}). In the simplest argument, the gaussian
fluctuations of the entropy translate into gaussian fluctuations in
the barrier, where the relative deviations of the two quantities, from
the most probable value, are given by \cite{XWbeta}:
\begin{equation} \label{relf} \delta \wF \equiv \frac{\delta
    F}{F_\smp} = \frac{\delta s_c}{s_c} = \frac{1}{2 \sqrt{D}},
\end{equation}
where $D$ is the liquid's fragility from Eq.(\ref{tau}). The quantity
$1/2\sqrt{D}$ varies between 0.05 and 0.25 or so, for known
glassformers, the low and high limits corresponding to strong and
fragile substances respectively.

\begin{figure}[t]
  \includegraphics[width=.6\columnwidth]{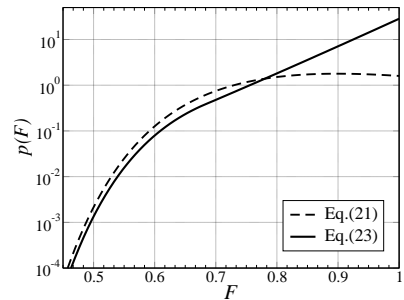} 
  \caption{\label{pFfig} The barrier distributions from Eqs.(\ref{pF})
    and (\ref{pFm}). The $\delta$-function portion of the distribution
    from Eq.(\ref{pF}) is not shown. $\delta \wF = 0.25$.}
\end{figure}

\begin{figure}[t]
\includegraphics[width=.85\columnwidth]{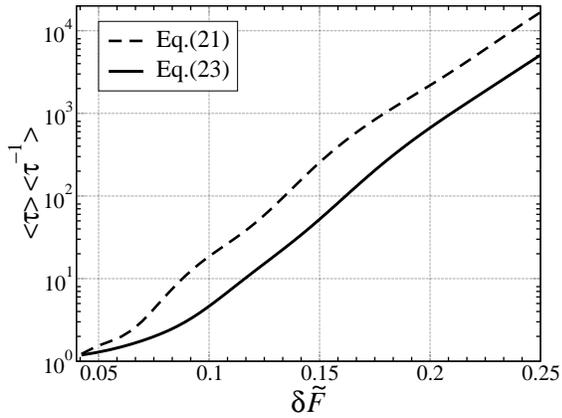}
\caption{\label{t_tt} The decoupling between the viscosity and the
  intrinsic ionic conductivity, as a function of the relative barrier
  width $\delta \wF$ from Eq.(\ref{relf}), at $T_g$. The dashed and
  solid line pertain to the specific barrier distributions from
  Eq.(\ref{pF}) and (\ref{pFm}) respectively.  }
\end{figure}

Xia and Wolynes (XW) further argued that the real barrier distribution
should be cut-off at the most probable value because a liquid region
with relatively low density of states is likely neighbors with a
relatively fast region \cite{XWbeta}. In addition we may recall that
in the library construction, the most likely liquid state is the one
where the liquid is {\em guaranteed} to have an escape trajectory
\cite{LW_aging}. This means that the most probable barrier is also the
maximum barrier. One may conclude then that the naive Gaussian
distribution is adequate at small barriers, but significantly
overestimates the probability of barriers larger than the typical
barrier.  Put another way, the trajectories corresponding to higher
than most probable barrier in the naive Gaussian, all contribute to
the $F \le F_\smp$ range. XW have implemented this notion by replacing
the r.h.s. of the simplest Gaussian distribution by a delta-function
centered at $F_\smp$ \cite{XWbeta}:
\begin{equation} \label{pF} p_1(\wF) = \frac{e^{-(1/\wF-1)^2/2\delta
      \wF^2}}{\sqrt{2 \pi (\delta \wF)^2} \wF^2} +
  \frac{1}{2}\delta(\wF-1),
\end{equation}
where $\wF \equiv F/F_\smp < 1^+$, and we took advantage of the
temperature-independence of the relative width in Eq.(\ref{relf}).
This approximate form does not use adjustable parameters and
quantitatively accounts for the correlation between the fragility and
the stretching exponent $\beta$ \cite{XWbeta}, and the deviations from
the Stokes-Einstein relation.  The distribution in Eq.(\ref{pF}) is
shown in Fig.\ref{pFfig}.  The only difference of Eq.(\ref{pF}) with
the XW's form is that they used a purely gaussian form for $\wF < 1$,
whereas we follow their own suggestion and employ the more accurate $F
\propto 1/s_c$ (where $s_c$ is gaussianly distsributed of course). The
accurate evaluation of the left wing of the distribution is imperative
in estimating the average rate $\tau_\smicro^{-1} e^{-F/k_B T}$,
because the latter is a rapidly varying function of $F$. ($k_B T$ is
significantly less that $\delta F$ at low temperatures.) Note that
because of the rapid decay of the exponential at small $\wF$ in
Eq.(\ref{pF}), accounting for the lowest order, quadratic fluctuations
of entropy suffices. The quantity $\la \tau \ra \la \tau^{-1} \ra$,
that characterizes the apparent decoupling, computed with the XW's
distribution, is shown with the dashed line in Fig.\ref{t_tt}, at
$T_g$, as a function of the relative distribution width $\delta \wF$.

How robust is the prediction based on the simple functional form for
the barrier distribution from Eq.(\ref{pF})? In spite of its
quantitative successes, one may argue that the true barrier
distribution should be a smoother function, near $\wF =1$.  One way to
see this is to computing $\epsilon''(\omega)$ from Eq.(\ref{epsmu})
via the distribution in Eq.(\ref{pF}): The obtained curves are a sum
of two peaks, one of which is broader, one the other is narrower than
the experimental $\epsilon''(\omega)$.  The two peaks correspond to
the half-Gaussian and the delta-function in Eq.(\ref{pF})
respectively. Let us see that knowing the precise form of the barrier
distribution however is not essential in quantitative estimates of the
decoupling so long as we account correctly for the overall width of
the distribution and its decay at the low barrier side.

It is straightforward to show that there exists a distribution that
(a) satisfies these requirements without introducing adjustable
constants, (b) reproduces the experimental $\epsilon(\omega)$ and does
as well as the XW form for the $\beta$ vs. $D$ correlation.  As we
have already discussed, the low barrier wing of the distribution in
Eq.(\ref{pF}) is adequate.  On the other hand, the high barrer wing
should include the contributions from both sides of the original
Gaussian peak, which are both of width $\delta F/2$. ``Stacking''
these two on top of each other, to the left of $F_\smp$, results in a
distribution of width $\delta F/4$ (see aslo Appendix).  Further,
based on the known $\epsilon(\omega)$ data, the barrier distribution
should be well approximated by an exponential, suggesting we use $p(F)
\propto e^{F/(\delta F/4)}$ near $F_\smp$. Indeed, this implies
$p(\tau) \propto \tau^{(4 k_B T/\delta F)-1}$.  At frequencies not too
close to the maximum of $\epsilon''(\omega)$ and the rapid drop-off at
small $\wF$, one has an approximate power law:
\begin{equation} \label{appr} \epsilon''(\omega) \simeq \int_0^\infty
  d\tau \, \tau^{(4 k_B T/\delta F)-1} \frac{\omega \tau}{1+(\omega
    \tau)^2} \propto \omega^{-4 k_B T/\delta F}.
\end{equation}
We thus arrive at the following form:
\begin{equation} \label{pFm} p(\wF) = \left\{ \begin{array}{ll}
      \frac{c_1}{\wF^2} e^{-(1/\wF-1)^2/2\delta \wF^2}, & \wF \le \wF_e\\
      \frac{c_2}{\wF^2} e^{\wF/(\delta \wF/4)}, & \wF_e < \wF \le 1,
    \end{array}
  \right.
\end{equation}
where $F_e$ and the normalization constants $c_1$ and $c_2$ are chosen
so that the distribution is normalized, continuous, and its first
derivative is continuous too. The distribution from Eq.(\ref{pFm}) is
plotted in Fig.\ref{pFfig}. The decoupling strength $\la \tau \ra \la
\tau^{-1} \ra$, computed for the composite distribution from
Eq.(\ref{pFm}), is shown in Fig.\ref{t_tt}, as a function of the
relative distribution width $\delta \wF$, at $T_g$.  We therefore
observe that in fragile liquids, the apparent time-scale separation
may reach as much as four orders of magnitude near the glass
transition - even though only one process is present! - because the
inrinsic ionic conductivity is dominated by the fastest relaxing
regions.

Conversely, when the apparent decoupling exceeds the intrinsic value
prescribed by Fig.\ref{t_tt}, we may conclude that ionic conduction
does not in fact require structural relaxation. This notion is of
significance for the mechanisms of electrical conductance in glasses
and will be discussed in detail in the Conclusions.

\begin{figure}[t]
\includegraphics[width=.85\columnwidth]{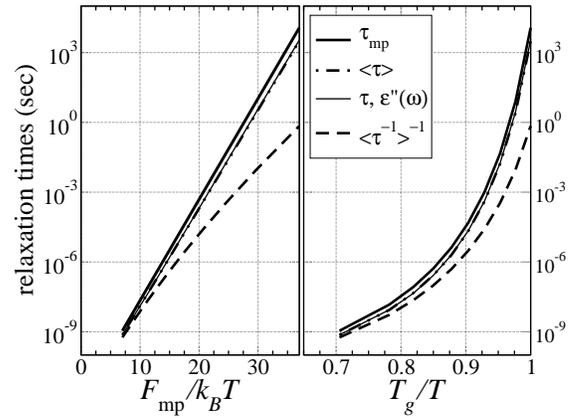}
\caption{\label{t_FT} Different relaxation times derived from the
  barrier distribution in Eq.(\ref{pFm}), as functions of the most
  probable barrier (left) and the corresponding temperature
  (right). $\delta \wF = 0.25$, corresponding to fragility $D=4$.}
\end{figure}

One may also illustrate the effects of apparent decoupling for a
specific value of fragility, by plotting several varieties of
relaxation times, as functions of the most probable barrier, or the
corresponding temperature, see Fig.\ref{t_FT}. (Given the time scale
at the glass transition, say $D T_K/(T_g - T_K) = \ln(10^{16}) \simeq
37$ (see Eq.(\ref{tau})), there is a one-to-one correspondence between
the fragility $D$ and the $T_g/T_K$ ratio.) We observe that the
average relaxation time and the one derived from the inverse of the
maximum position of $\epsilon''(\omega)$ are close, and are near the
most probable value of the relaxation time. (The $\epsilon''(\omega)$
was computed using Eqs.(\ref{epsfull}) and (\ref{pFm}), see below.)
The apparent conductivity relaxation time is strongly decoupled,
consistent with data of Howell at el. \cite{Howell1974} for CKN. Note
that the value of fragility used in Fig.\ref{t_FT}, $D=2$, is probably
smaller than in CKN. In addition, we have ignored here, for clarity,
the effects of barrier softening \cite{LW_soft}, that would require
introducing a system-specific adjustable constant $T_A$.  The latter
effects would change the slopes of the curves somewhat, without
affecting their vertical separations.

To test the predictions from Figs.\ref{t_tt} and \ref{t_FT}, one needs
to know the width of the barrier distribution for $\alpha$-relaxation.
As already mentioned, the gross features of this distribution have
been predicted by the RFOT theory, and have lead to quantitative
predictions of the correlation between the stretching exponent $\beta$
and the fragility $D$, and the deviations from the Stokes-Einstein
relation. The corresponding trends are as follows: more fragile
liquids are predicted to have broader barrier distribution leading to
a smaller value of $\beta$, and vice versa for stronger substances
\cite{XWbeta}.  A correlation with the fragility comes about by virtue
of Eq.(\ref{relf}). Several {\em \`{a} priori} ways to determine
$\beta$ and $D$ have been employed, by experimenters, that sometimes
produce conflicting results.  For example, the fragility extracted
from $\tau_\sigma$ will be consistently lower than that extracted from
the mechanical relaxation time $\tau_s$, because $\tau_\sigma <
\tau_s$. The exponent $\beta$ from the stretched exponential is
extracted from fits of various relaxation processes to a stretched
exponential profile $e^{-(t/\tau)^\beta}$.  Alternatively, one may
choose to fit the Fourier transform of the stretch exponential, or the
Cole-Davidson form, to the imaginary part of $\epsilon(\omega)$ in
insulators \cite{LindseyPatterson}. These usually produce comparable
results for the corresponding exponent $\beta$, with a notable
exception of ionic conductors, which happen to be the main focus of
this paper. In ionically conducting systems, the dc component of the
full dielectric response from Eq.(\ref{epsfull}) largely ``swamps''
the ac part so that reliable determinations of the latter are
complicated. The reader is reminded that dielectric measurements on
ion melts are difficult because electrodes generally block ionic
current. The effects of build-up charge are often treated
phenomenologically, by means of equivalent circuits \cite{Macdonald,
  Pimenov}. Given these complications, many have chosen to plot the
reciprocal of $\epsilon(\omega)$, i.e. the dielectric modulus
\cite{Howell1974, Macdonald}:
\begin{equation} \label{Mo} M(\omega) \equiv 1/\epsilon(\omega).
\end{equation}
$M(\omega)$ is well behaved and even shows a peak in the imaginary
component, similarly to $\epsilon(\omega)$ of a near insulator.  In
the absence of an {\em \`{a} priori} microscopic picture and by
analogy with $\epsilon(\omega)$, one might interpret this peak as as
the response of the electric field $\bE$ to the dielectric
displacement $\bD$. This in fact would be appropriate in a layered
dielectric \cite{Howell1974}. See also the discussions in
Refs.\cite{Elliott1994, Roling1998, ColeTombari, Sidebottom1995,
  Moynihan1994, Dyre1991, Doi1988}. Yet the resulting values of the
most probable relaxation time and the stretching exponent deviate from
those obtained with other methods \cite{Angell, Sidebottom1995,
  Pimenov}. In fact, the modulus-derived $\beta_M$ increases, while
the width of the $M''(\omega)$ peak decreases with lowering the
temperature, in conflict with the general trends for poor conductors,
and the conclusions of the RFOT theory.

\begin{figure}[t]
\includegraphics[width=.9\columnwidth]{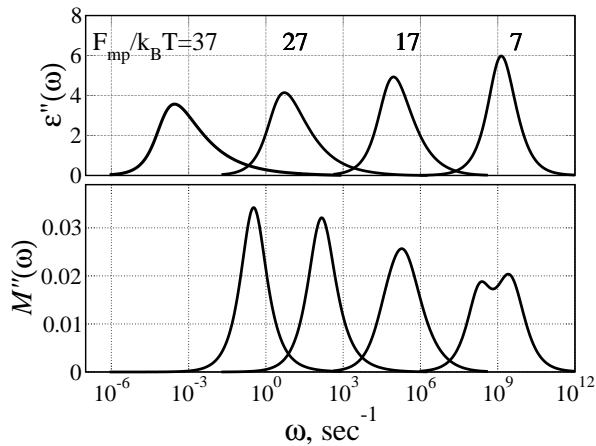}
\caption{\label{epsM} The top panel shows the imaginary part of the
  non-conductive part of the dielectric response, $\epsilon(\omega)$,
  Eq.(\ref{epsmu}), for four values of the most probable
  $\alpha$-relaxation barrier: $F_\smp/k_B T =
  \ln(\tau/\tau_\smicro)$, as indicated on the graph. ($\delta \wF =
  0.25$, $\tau_\smicro = 1$ ps.) These barrier values were chosen
  because for $\tau/\tau_\smicro > e^{7} \simeq 10^{3}$, the RFOT
  approach is quantitatively accurate, while $\tau/\tau_\smicro \simeq
  e^{37} \simeq 10^{16}$ is at the upper limit of the dynamical range
  routinely accessible in the lab. The bottom panel show the
  corresponding modulus, from Eq.(\ref{Mo}), that includs the dc-part,
  due to the intrinsic ionic current. The same four values of the
  barrier are used. We have used the CKN's values for $\epsilon_0$ and
  $\epsilon_\infty$, and assumed that $\frac{\Delta \mu_\smol^2}{ a^3
    k_B T} \simeq 2 (\epsilon'_0-\epsilon_\infty)$, see the discussion
  preceeding Eq.(\ref{Maxwell_cond}).}
\end{figure}

The RFOT theory and the present results allow one to address these
difficulties, to which we devote the rest of this Section. One first
notes that structural reconfigurations are {\em compact}, and so the
layered-dielectric view of supercoold melts is not microscopically
justified. We next plot, in the top panel of Fig.\ref{epsM}, the
non-conductive $\epsilon_\sins''(\omega)$, from Eq.(\ref{epsmu})
averaged with respect to the barrier distribution from Eq.(\ref{pFm}).
We have used CKN's values for $\epsilon_0$ and $\epsilon_\infty$, as
before.  For the sake of argument, we use $\delta \wF = 0.25$,
corresponding to $\beta \simeq 0.4$ at $T_g$.  In Fig.\ref{epsM},
bottom, we show the imaginary part $M''(\omega)$, of the full modulus.
Clearly the two functions exhibit qualitatively different behaviors.
Note that the effect of the dc component on the apparent relaxation
profile has been discussed previously \cite{Johari1988, Roling1998},
including the possibility of a double peak \cite{Johari1988}. The
latter has been observed by Funke at el. \cite{Funke}, but has not
been reproduced by Pimenov at el. \cite{Pimenov}. Nevertheless, the
dielectric modulus obtained here is qualitatively consistent with
CKN's data from Ref.\cite{Pimenov}.  Finally note that for smaller
dc-conductivities, the modulus data would become more similar to
$\epsilon''(\omega)$.

\begin{figure}[t]
  \includegraphics[width=.7\columnwidth]{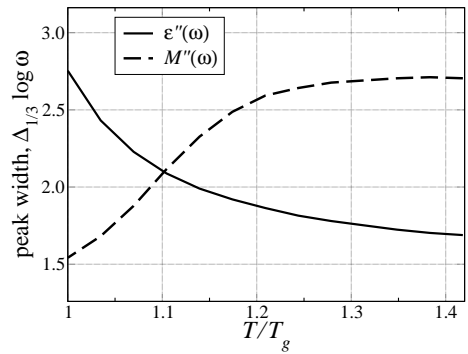} 
  \caption{\label{w_T} The widths of the peaks in the imaginary parts
    of the dielectric constant $\epsilon(\omega)$ and modulus
    $M(\omega)$, as functions of temperature
    (c.f. Fig.\ref{epsM}). $\delta \wF = 0.25$.}
\end{figure}

We conclude from the above analysis that if one were to use the
modulus data to extract the characteristics of the barrier
distribution, one must measure first the dc-current, add it to the
$\epsilon_\sins(\omega)$ from a microscopic theory, and then compare
the result to the measured $M(\omega)$ data. But again, because of the
large contribution of the dc component, the corresponding fits would
not discriminate well between different forms of
$\epsilon_\sins(\omega)$. On the other hand, treating the electric
field as a response to the displacement may lead to erroneous
conclusions on the temperature dependence of the barrier width. In
fact, the barrier widths derived from $\epsilon_\sins''(\omega)$ or
$M''(\omega)$ show the opposite trends, as we have seen already in
Fig.\ref{epsM}. One may further quantify this observation: In the
absence of a microscopic theory, one often characterizes the width of
the $\epsilon_\sins''(\omega)$ peak by a stretching exponent $\beta$,
as derived e.g. from Davidson-Cole fits. The distribution from
Eq.(\ref{pFm}) indeed gives rise a power law behavior, consistent with
Eq.(\ref{appr}), see Appendix. In contrast, the corresponding
$M''(\omega)$ curves do not exhibit a similar power-law behavior.  I
have chosen to illustrate the opposite temperature trends in the
widths of $\epsilon_\sins''[\ln(\omega)]$ and $M''[\ln(\omega)]$
peaks, by measuring the latter widths at one-third-height and plotting
them as functions of temperature, see Fig.\ref{w_T}.  Similar opposite
trends, too, would be observed for the corresponding {\em apparent}
barrier widths or effective $\beta$'s. Clearly, interpreting the
dielectric modulus of an ionic conductor as a response function may
lead to a significant underestimation of the actual barrier width at
low temperatures, and qualitatively incorrect conclusions on the
temperature dependence of the width.

\section{Conclusions}

We have computed, from the first principles, the viscosity and the
intrinsic ionic conductivity of supercooled liquids. The viscosity is
determined by four microscopically defined quantities: the length
scale of the local chemical order that sets in at temperature
$T_\scr$, where liquid dynamics become activated; the Lindemann
length, characterizing molecular displacements at the mechanical
stability edge; the temperature; and the average relaxation time
$\tau$ of the activated reconfigurations that dominate the liquid
dynamics below $T_\scr$. The extraordinarily long $\tau$ range is what
gives rise to the high viscosity of the liquid when it approaches the
glass transition. When the local chemically stable units (or
``beads'') are charged, the fluid will also exhibit an ionic
conductivity, which we have called the ``intrinsic'' conductivity, to
constrast it with electric conduction via delocalized electronic
carriers or via mobile ions that are not bonded to the metastable
aperiodic lattice forming the supercooled liquid. Computing the
conductivity requires an additional microscopic characteristic, the
electric charge on a ``bead''. Fortunately, this additional parameter
may be deduced from the ac dielectric response, which we have also
estimated. Perhaps the main finding of this work is that in contrast
with the viscosity, the ionic conductivity is dominated by the fastest
relaxing regions in the liquid, as reflected in Eq.(\ref{sij}).

We have discussed ways to test the above predictions, the most
important aspect of which is the large separation, or ``decoupling'',
between the apparent time scales, suggested previously by viscosity
and ionic conductivity data on purely phenomenological grounds. We
have shown that such apparent time-scale separation is indeed expected
because of the very broad barrier distribution for
$\alpha$-relaxation, derived earlier in the Random First Order
Transition (RFOT) methodology. The decoupling thus stems essentially
from the same cause as the violation of Stokes-Einstein relation in
supercooled liquids \cite{XWhydro}. Now, we have seen that the value
of the decoupling is not very sensitive to the precise form of the
barrier distribution so long as one acounts for the RFOT-derived gross
characteristics of this.  We have thus quantified the degree of
``decoupling'': The intrinsic ionic conductivity was argued to
decouple at most by four orders of magnitude from the low-frequency
momentum transport. Conversely, any conductivity exceeding this limit
must be due to other charge carriers that do not disturb the liquid
structure beyond typical vibrational displacements. Indeed, suppose
the apparent decoupling exceeds the value prescribed by the width of
the barrier distribution. This means that there will be ions that
travel a distance exceeding the Lindemann length in a time it takes
the local environment to relax. Therefore, local relaxation is not a
necessary condition for a non-zero current of these ions. Some
interaction with relaxation may still be present, however at large
enough decouplings, we may say that the ion (or any other carrier)
interacts with the liquid as if the latter were a perfectly stable,
albeit disordered lattice.  In such cases, one may think of the ionic
current in superionic conductors in terms of regular, not slaved
diffusion.  In regular diffusion, the total travel time is dominated
by the slowest step, in contrast to Eq.(\ref{sij}).

The intrinsic difficulty in experimental assessment of the barrier
distribution in moderately conductive melts is that the dc current
dominates the overall dielectric response. This gives rise to
ambiguities as to what the actual width of the barrier distribution
is, since mechanical relaxation and dielectric {\em modulus} data
disagree.  We have shown that this is expected, and argued that the
mechanical relaxation offers the preferred method of estimating the
actual barrier distribution.

{\em Acknowledgments:} The author thanks Peter G. Wolynes for critical
comments and useful suggestions. He gratefully acknowledges the GEAR,
the New Faculty Grant, and the Small Grant Programs at the University
of Houston.

\section*{Appendix}

\begin{figure}[t]
  \includegraphics[width=.95\columnwidth]{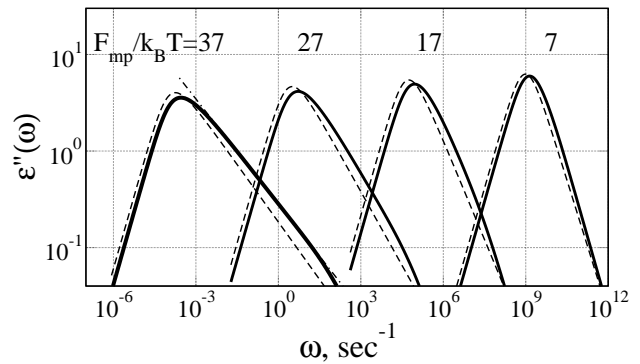} \vspace{3mm}
  \caption{\label{epsom} The four thick lines are the same as in
    Fig.\ref{epsM}, but in the double-log scale. The thin lines are
    the Davidson-Cole forms, following from a simple approximation,
    see Appendix. The dash-dotted line illustrates how the stretching
    exponent $\beta$ was extracted from the curves.}
\end{figure}

\begin{figure}[t]
\includegraphics[width=.85\columnwidth]{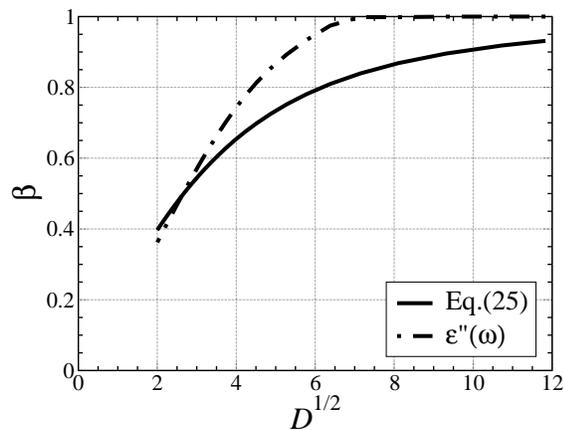}
\caption{\label{bD} Two approximate relations of the stretching
  exponent $\beta$ to the fragility $D$ from the Vogel-Fulcher form,
  from Eq.(\ref{betaD}) and as derived from the slopes of the
  high-frequency wing of the $\epsilon''(\omega)$ peaks, such as in
  Fig.\ref{epsom}.}
\end{figure} 

Let us see that the distribution in Eq.(\ref{pFm}) is qualitatively
consistent with experimental $\epsilon_\sins (\omega)$ and the
empirical correlation of $\beta$ and $D$.  For this, we replot the top
panel of Fig.\ref{epsM} in the double-log format, in
Fig.\ref{epsom}. We note the general adequacy of the barrier
distribution from Eq.(\ref{pFm}): Similarly to the experimental
$\epsilon(\omega)$ in poor conductors, the resulting high-frequency
wing is significantly broader than the low-frequency one. Note that
the actual data would also often display an {\em additional}
high-frequency wing, which is ascribed to the secondary,
$\beta$-processes, also called Johari-Goldstein relaxation
\cite{JohariGoldstein}. (See \cite{LunkLoidl} for a review). The
present results suggest that $\beta$-relaxation does not contribute to
the intrinsic ionic conductivity. At any rate, the derived
$\epsilon(\omega)$ show several decades of nearly power-law decay,
allowing one to extract the corresponding exponent: $\epsilon(\omega)
\propto \omega^{-\beta}$.  The effective $\beta$'s were deduced from
the slopes of the curves at the points of maximum second derivative,
as exemplified by the dash-dotted line in Fig.\ref{epsom}. The
dependence of the thus obtained exponent $\beta$ on the fragility $D$,
at a fixed $F_\smp/k_B T = 37$, is shown by the dashed-dotted line in
Fig.\ref{bD}. This $\beta$ is, again, qualitatively consistent with
experiment. Greater accuracy should not be expected here, as we have
not treated the higher-frequency range associated with
$\beta$-relaxation, which would affect the experimentally determined
stretching exponents.

In addition, we verify that the informal argument in the main text
that the width of the barrier distribution should be about $\delta F/4
$, at the half-height or so.  Indeed, for a gaussian barrier
distribution with width $\delta F/4 = F/8\sqrt{D}$ implies the
following approximate expression for the stretching exponent $\beta$
at $T_g$ (c.f. Eq.(9) of Ref.\cite{XWbeta}):
\begin{equation} \label{betaD} \beta = \left[ 1 + \left(
      \frac{F/k_BT}{8 \sqrt{D}} \right)^2 \right]^{-1/2},
\end{equation}
shown as the solid line in Fig.\ref{bD}. This expression is in very
good agreement with experiment, see Fig.2 from Ref.\cite{XWbeta}.  (At
$T_g$ on scale $\tau/\tau_\smicro = 10^{16}$, $F/k_B T \simeq 37$.)
Note also Eq.(\ref{betaD}) is consistent with Eq.(\ref{relf}),
assuming the Davidson-Cole \cite{CD} and William-Watts \cite{KWW}
stretching exponents $\beta$ are close \cite{LindseyPatterson}.  That
the latter is the case indeed I demonstrate by graphing the
Davidson-Cole (DC) form $(\epsilon_{\tDC}(\omega) - \epsilon_\infty) =
(\epsilon_0 - \epsilon_\infty) (1 - i \omega \tau)^{-\beta}$ \cite{CD}
with $\tau = \tau_\smicro e^{F_\smp/k_B T}$ and $\beta$ from
Eq.(\ref{betaD}). These are shown in Fig.\ref{epsom} as thin dashed
lines.

\bibliography{/Users/vas/Documents/tex/ACP/lowT}

\end{document}